\documentclass[12pt,preprint]{aastex}

\newcommand{\vlsr}  {$v_{\rm LSR}$}
\newcommand{\kms}   {km~s$^{-1}$}
\newcommand{\metha} {CH$_3$OH} 
\newcommand{\J}[2]  {\mbox{$J$=#1$\to$#2}}

\received{2001 December 14}
\begin{document}

\accepted{}


\title{IRAS 21391+5802: The Molecular Outflow and its Exciting Source}

\author{Maria T. Beltr\'an\altaffilmark{1, 2}, Jos\'e M.\
Girart\altaffilmark{2}, Robert
Estalella\altaffilmark{2}, Paul T. P. Ho\altaffilmark{1}, and Aina
Palau\altaffilmark{2}}

\altaffiltext{1}{
Harvard-Smithsonian Center for Astrophysics,
60 Garden Street, Cambridge, MA 02138;
{\tt{mbeltran@cfa.harvard.edu}, \tt{pho@cfa.harvard.edu}}
}
\altaffiltext{2}{
Departament d'Astronomia i Meteorologia, Universitat de Barcelona,
Av.\ Diagonal 647, E-08028 Barcelona, Catalunya, Spain;
{\tt{jgirart@am.ub.es}, \tt{robert.estalella@am.ub.es},
\tt{apalau@am.ub.es}}
}

\begin{abstract}

We present centimeter and millimeter observations of gas and dust around
IRAS~21391+5802, an intermediate-mass source embedded in the core of IC~1396N.
Continuum observations from 3.6~cm to 1.2~mm are used to study the embedded
objects and overall distribution of the dust, while molecular line
observations of CO, CS, and CH$_3$OH are used to probe the structure and
chemistry of the outflows in the region. The continuum emission at centimeter
and millimeter wavelengths has been resolved into three sources separated
$\sim 15\arcsec$ from each other, and with one of them, BIMA~2, associated
with IRAS~21391+5802. The dust emission around this source shows a very
extended envelope, which accounts for most of the circumstellar mass of
$5.1~M_\sun$. This source is powering a strong molecular outflow, elongated in
the E--W direction, which presents a complex structure and kinematics. While
at high outflow velocities the outflow is clearly bipolar, at low outflow
velocities the blueshifted and redshifted emission are highly overlapping, and
the strongest emission shows a V-shaped morphology. The outflow as traced by
CS and CH$_3$OH exhibits two well differentiated and clumpy lobes, with two
prominent northern blueshifted and redshifted clumps. The curved shape of the
clumps and the spectral shape at these positions are consistent with shocked
material. In addition, CS and CH$_3$OH are strongly enhanced toward these
positions with respect to typical quiescent material abundances in other 
star-forming regions. These kinematical and chemical evidences suggest that
the clumps are tracing gas entrained within the surface of interaction between
the molecular outflow and the dense ambient quiescent core, and that the
morphology of the molecular outflow is a result of this interaction. The
circumstellar mass together with the power-law index of the dust emissivity
measured, $\beta=1.1\pm0.3$, and the fact that the source is driving a
molecular outflow are consistent with the source BIMA~2 being an embedded
intermediate-mass protostar. In addition, the source fits very well
correlations between source and outflow properties found for low-mass Class~0
objects. The other two sources in the region, BIMA~1 and BIMA~3 have a mass of
$0.07~M_\sun$, and their dust emissivity index, $\beta<0.3$ and
$\beta=0.1\pm0.3$, respectively, is consistent with more evolved objects.
BIMA~1 is also driving a very collimated and small bipolar outflow elongated
in the N--S direction.

\end{abstract}


\section{Introduction}

It is well known that Young Stellar Objects (YSOs) are embedded in dust and
gas. For low-mass stars, circumstellar dusty material can be found in two
basic structures: envelopes and disks. Theory outlines an evolutionary
scenario in which these structures evolve from envelopes to disks as the
central object evolves (Adams, Lada, \& Shu 1987; Shu, Adams, \& Lizano 1987)
from the youngest Class~0 objects, surrounded by infalling envelopes, to the
optically visible T~Tauri stars, with circumstellar disks. Looney, Mundy, \&
Welch (2000) have carried out sub-arcsecond millimeter wavelength observations
of the dust emission around low-mass objects, and have shown that the more
embedded objects have most of their emission in large scale structures,
consistent with envelopes, while optical/IR objects have compact continuum
emission, associated with disks, with little or no extended emission. Molecular
line observations also show a tendency of the gas emission to be stronger
toward protostar candidates than toward T~Tauri stars (e.g.\ Ohashi et al.\
1991, 1996; Chen, Fukui, \& Yang 1992). 

This well-defined evolutionary sequence found for low-mass stars, however,
does not have a counterpart for higher mass stars, such as Herbig~Ae and Be
stars and their intermediate-mass precursors with masses in the range
$2~M_\sun\leq M_\star\leq10~M_\sun$. Circumstellar disks, similar to those
surrounding T~Tauri stars, have been detected around Herbig~Ae stars
(Guilloteau \& Dutrey 1998; Mannings \& Sargent 1997, 2000; Natta et al.\
2001), but not around Herbig~Be stars (Natta, Grinin, \& Mannings 2000; Fuente
et al.\ 2001). The non-detection of circumstellar disks around Herbig~Be stars
has been interpreted as a consequence of the rapid evolution of the
circumstellar structures and of the outflows as compared to T~Tauri and
Herbig~Ae stars (Natta et al.\ 2000; Fuente et al.\ 2001). Regarding the
intermediate-mass very YSOs, the situation is even less studied and less well
understood due to the difficulty of identifying the youngest truly protostars,
which are still in the infalling envelope phase. In addition, these objects
are rare in comparison with their low-mass counterparts and tend to be located
at greater distances. The immediate vicinity of intermediate-mass protostars
is a very complex environment, where the extended emission is usually resolved
into more than one source when observed at high resolution (e.g.\
G173.58+2.45: Shepherd \& Watson 2002). In addition, the molecular outflows
driven by intermediate-mass objects are more energetic. Thus, their
interaction with the circumstellar gas and dust material surrounding the
protostars is expected to be stronger and more dramatic, disrupting the
envelopes and pushing away the dense gas at high velocities. Although these
outflows are, in general, less collimated and more chaotic than those of
low-mass stars (e.g.\ NGC~7129: Fuente et al.\ 2001), this could probably be
due to observational constraints. Since these regions are located, on average,
further away, these outflows are usually observed with less linear spatial
resolution than nearby low-mass flows. However, when observed with high
angular resolution their appearance seems to be more collimated and less
chaotic (e.g.\ HH~288: Gueth, Schilke, \& McCaughrean 2001). Therefore, there
is little doubt that the study of intermediate-mass star-forming regions is a
very complicated task, but attractive at the same time, due to all the
phenomena that are taking place in such environments. It is also clear that in
order to understand the heterogeneity of the more evolved Herbig~Ae and Be
stars, it is necessary to study such objects at a very early stage of their
evolution, when both envelopes and circumstellar disks exist around them.  

In an attempt to build our understanding of intermediate-mass stars and compare
their morphology and evolution with those of low-mass stars, we carried out
interferometric radio observations of the dust and gas around IRAS~21391+5802,
a very young intermediate-mass object. IRAS~21391+5802 is a cold far-infrared
source, with a bolometric luminosity estimated to range from 235~$L_\sun$
(Saraceno et al.\ 1996) to 440~$L_\sun$ (Sugitani et al.\ 2000). It is deeply
embedded in the bright-rimmed globule IC~1396N, which is located at a distance
of 750~pc (Matthews 1979). This source exhibits all of the signposts of an
extremely young object, such as strong submillimeter and millimeter continuum
dust emission (Wilking et al.\ 1993; Correia 2000; Sugitani et al.\ 2000;
Codella et al.\ 2001), high-density gas (Serabyn, G\"usten, \& Mundy 1993;
Cesaroni, Felli, \& Walmsley 1999; Codella et al.\ 2001), and water maser
emission (Felli, Palagi, \& Tofani 1992; Tofani et al.\ 1995; Patel et al.\
2000). Sugitani et al.\ (1989) discovered an extended CO bipolar outflow, which
has also been mapped by Codella et al.\ (2001). In addition, recent
near-infrared images by Nisini et al.\ (2001) have revealed the presence of a
collimated H$_2$ jet driven by IRAS~21391+5802.

\section{Observations}

\subsection{VLA observations}

The 3.6~cm continuum observations were carried out in 2000 September 23
using the Very Large Array (VLA) of the NRAO\footnote{The National Radio
Astronomy Observatory is a facility of the National Science Foundation
operated under cooperative agreement by Associated Universities, Inc.} in
the D configuration. The field was observed for a total of about 1~hr
45~m, and the phase center was set to the position $\alpha({\rm
J2000})=21^{\rm h}40^{\rm m}41\fs79$, $\delta({\rm
J2000})=58\degr16\arcmin12\farcs4$, which coincides with the 3.1~mm
position of one of the sources in the field, BIMA~2 (see
$\S~\ref{bima_res}$). Absolute flux calibration was achieved by observing
3C~286, with an adopted flux density of 5.18~Jy. Phase was calibrated by
observing 2148+611, which has a bootstrapped flux of $0.735\pm0.007$~Jy.
CLEANed maps were made using the task IMAGR of the Astronomical Image
Processing System (AIPS) software of NRAO, with the ROBUST parameter of
Briggs (1995) set equal to 5, which is close to natural weighting. The
resulting synthesized beam was $17\farcs9\times8\farcs8$ at P.A.\ $=
-80$\degr, and the rms noise of the maps was 20~$\mu$Jy\,beam$^{-1}$.
For the maps made with uniform weighting the synthesized beam was
$12\farcs6\times5\farcs9$ at P.A.\ $=-82$\degr, and the rms noise was
slightly higher, 32~$\mu$Jy\,beam$^{-1}$.

\subsection{BIMA observations}

Millimeter interferometric observations were carried out with the BIMA
array\footnote{The BIMA array is operated by the Berkeley-Illinois-Maryland
Association with support from the National Science Foundation.} in the B and
the C configurations between March 2000 and January 2001. The digital
correlator was configured to observe simultaneously the continuum emission and
some molecular lines. Details of the observations are given in
Table~\ref{tbima}. The phase center was located at $\alpha({\rm
J2000})=21^{\rm h}40^{\rm m}42\fs40$, $\delta({\rm
J2000})=58\degr16\arcmin10\farcs0$,  the position of IRAS~21391+5802.
Amplitude and phase were calibrated by observations of BL~Lac, whose flux
density was determined relative to Uranus. The uncertainty in the amplitude
calibration is estimated to be $\sim 20$\%. The data were calibrated and
analyzed using standard procedures in the Multichannel Image Reconstruction
Image Analysis and Display (MIRIAD) software package (Sault, Teuben, \&
Wright 1995). The quality of the continuum data at 2.7~mm was not good enough
to use the data for mapping and analysis.

\section{Results}
\subsection{Centimeter continuum emission}

In Figure~1 we show the natural-weight map at 3.6 cm superposed on the red
image of the Digitized Sky Survey\footnote{The Digitized Sky Survey was
produced at the Space Telescope Science Institute under U.S. Government
grant NAG W-2166. The images of these surveys are based on photographic
data obtained using the Oschin Schmidt Telescope on Palomar Mountain and
the UK Schmidt Telescope. The plates were processed into the present
compressed digital form with the permission of these institutions.} 2. The
extended 3.6 cm emission matches the optical emission of a bright-rimmed
cloud, catalogued as BRC 38 by Sugitani, Fukui, \& Ogura (1991) and
belonging to the HII region IC~1396N. Above the 5 $\sigma$ detection
threshold, in addition to the extended emission, we detected 5 sources,
VLA 1 to 5. The first three, unresolved in the natural-weight map, appear
resolved into three sources separated $\sim 15''$ from each other in the
uniform map, shown in Figure~2. The central positions and flux densities
of the detected sources are given in Table~\ref{tvla}.  The spectral
indices between 21 and 3.6 cm were obtained using the NRAO VLA Sky Survey
(Condon et al.\ 1998) to measure the 21 cm flux densities, which provides an
angular resolution of $\sim 45\arcsec$. The results are
shown in Table~\ref{tvla}.

The spectral index of BRC 38 is $-0.03$, characteristic of optically thin
free-free emission of a HII region. VLA~4 has a spectral index of $-0.83$,
which is consistent with non-thermal emission and is probably a background
source.  The emission of VLA~5, with a measured spectral index of 0.14, may
have a thermal component or is time variable (and non-thermal), giving an
apparently positive spectral index. As seen in Figure~1, VLA~5 is close to an
optical star of spectral type F identified as Cl Trumpler 37 1030 by Alknis
(1958). However, the offset of $\sim15''$ between the radio continuum emission
and the star is too large, and it suggests that they are likely not associated.
VLA~1, VLA~2 and VLA~3 coincide with the sources BIMA~1, BIMA~2 and BIMA~3
respectively, detected at mm wavelengths with BIMA (see next section). VLA 2
lies inside the error ellipse of IRAS 21391+5802 and coincides with the near
infrared source IRS~2 (Wilking et al.\ 1993). VLA~3 also lies within the error
ellipse of the IRAS source, and it is the strongest of the three sources. Water
masers have been found associated with both sources (Tofani et al.\ 1995)

\subsection{Dust emission}
\label{bima_res}

The millimeter continuum emission around the embedded IRAS~2139+5802 was imaged
at 3.1 and 1.2~mm with the BIMA, and the maps are shown in Figure~3. As can be
seen in this figure, the continuum emission is dominated by an extended clumpy
structure, especially at 3.1~mm, which splits into two main peaks visible at
both wavelengths and separated $\sim 13\arcsec$ ($\sim 10^4$~AU at the distance
of the source), and a weaker peak toward the west, visible at 3.1~mm; hereafter
we will refer to these three clumps as BIMA~1, BIMA~2 and BIMA~3. Codella et
al.\ (2001) detected two unresolved continuum sources by using the OVRO
interferometer at millimeter wavelengths, called A and B, which are coincident
with BIMA~2 and BIMA~3, respectively. The dust emission has not been resolved
at submillimeter wavelengths by using the JCMT telescope (Correia 2000), which
provides a resolution of $\sim 8\arcsec$ at 450~$\mu$m, similar to the
resolution of our 3.1~mm map. However, the extended submillimeter emission
suggests the presence of multiple objects in the region. A possible explanation
for this could be that the interferometer is likely resolving out some of the
extended cold dust emission, and thus, it is more sensitive to more compact
structures. In fact, extrapolating the flux density of 10.5~Jy obtained at
850~$\mu$m by Correia (2000) to millimeter wavelengths by using typical
spectral indices in the range 3--4, it is clear that the interferometer
observations are indeed missing a significant amount of emission. Embedded
sources are very often members of multiple systems, as already noticed in
deeply embedded low-mass cores (Looney et al.\ 2000; Motte \& Andr\'e 2001),
which indicates that truly isolated star formation is rare. The positions and
flux densities of the millimeter sources at 3.1 and 1.2~mm are given in
Table~\ref{tmm}. At millimeter wavelengths the contribution by free-free
emission, which is expected to be dominant at centimeter wavelengths (e.g.\
Anglada 1995; Anglada et al.\ 1998; Beltr\'an et al.\ 2001) is likely small as
compared to the thermal dust emission. Wilking et al.\ (1989) observed the 
region at 2 cm with the VLA and did not detect any of the sources at a 
$4\sigma$ level of 0.51 mJy. By using this upper limit for the flux density at
2~cm, we determined the spectral index at centimeter wavelengths, $\alpha<1.8$ for
BIMA~1, $\alpha<1.3$ for BIMA~2, and $\alpha<0.4$ for BIMA~3. On the other
hand, Reynolds (1986) modeled the emission for an inhomogeneous source, for
which the free-free emission can come from both optically thin and optically
thick regions, and found that for a jet with constant velocity, temperature and
ionization fraction the spectral index should be always $<1.3$. This case is
the most likely scenario for outflow central sources (e.g.\ Anglada et al.\
1998; Beltr\'an et al.\ 2001). Thus, for BIMA~1 we assumed an upper limit for
the spectral index of 1.3. Extrapolating the fluxes obtained at 3.6~cm to
millimeter wavelengths with these spectral indices, we found that the expected
free-free emission at millimeter wavelengths is $\lesssim 5.0$~mJy ($<79$\% of
the millimeter emission; see Table~\ref{tmm}) for BIMA~1, $\lesssim 6.4$~mJy
($<10$\%) for BIMA~2, and $\lesssim 1.1$~mJy ($<8$\%) for BIMA~3. Thus, for
BIMA~2 and BIMA~3 the dust emission dominates at millimeter wavelengths, while
for BIMA~1 it is not clear from our results.

The strongest source detected at millimeter wavelengths in the region is
BIMA~2. As can be seen in the maps, BIMA~2 is the object associated with the
infrared source IRAS~21391+5802. Its millimeter emission clearly shows two
components, a centrally peaked source, plus an extended component, elongated
roughly in the N--S and the NE--SW direction at
3.2~mm and 1.3~mm, respectively. The comparison of the peak intensities
measured at both wavelengths with the total integrated flux densities (see
Table~\ref{tmm}) indicates that a considerable fraction of the flux density
($\gtrsim 50$\%) arises from the envelope. By fitting elliptical Gaussians to
the maps, we measured the deconvolved size of the source, which is
$\sim5500\times3700$~AU at P.A.\ $\simeq10\arcdeg$ at 3.1~mm, and
$\sim2000\times1600$~AU at P.A.\ $\simeq-21\arcdeg$ at 1.2~mm (see
Table~\ref{tmm}). Note that the deconvolved size is dependent on the
synthesized beam, indicating that the source does not have a perfect Gaussian 
shape structure. These sizes are consistent with the values found for envelopes
around low- and intermediate-mass protostars (e.g.\ Wilking et al.\ 1989;
Andr\'e, Ward-Thompson, \& Barsony 1993; Hogerheijde \& Sandell 2000; Looney et
al.\ 2000; Fuente et al.\ 2001). This source also exhibits centimeter emission
(see Table~\ref{tvla}), and it is driving a molecular outflow detected in the
region (see next section).

The source BIMA~3, located $\sim 13\arcsec$ ($\sim 9800$~AU in projection)
southeast of BIMA~2, has also been detected at both millimeter wavelengths. The
source is associated with centimeter emission (see Table~\ref{tvla}) and with
water maser emission (Tofani et al.\ 1995), indicating that BIMA~3 is a young
stellar object, too. However, its emission is more compact than the emission
around BIMA~2, and it has only been resolved at 1.2~mm (see Table~\ref{tmm}),
with a deconvolved size of $\sim1600\times1500$~AU at P.A.\ $\simeq-1\arcdeg$.
This deconvolved size is also very dependent on the angular resolution, having
a value of $\sim900\times300$~AU at P.A.\ $\simeq-64\arcdeg$ if we use only the
B configuration data. 

Finally, the other source detected in the region is BIMA~1. This source is a
faint unresolved clump only detected, as stated before, at 3.1~mm (see
Fig.~3). This source, also detected at centimeter wavelengths (see
Table~\ref{tvla}), is $\sim 16\arcsec$ ($\sim 12000$~AU) southwest of BIMA~2,
and $\sim 23\arcsec$ ($\sim 17300$~AU) northwest of BIMA~3, and it could be
the driving source of a very collimated N--S bipolar outflow discovered in the
region (see next section).

\subsection{The molecular outflow}

The molecular outflow associated with IRAS~21391+5802 has been previously
studied through lower resolution CO (Sugitani et al.\ 1989; Codella et al.\
2001), CS (Serabyn et al.\ 1993; Codella et al.\ 2001), SiO and HCO$^+$
observations (Codella et al.\ 2001). The observations reported in this study
improve significantly the angular resolution of the maps of the region: the
angular resolution of $\sim 4\arcsec$ for the CO is at least  6 times better
in beam area than previous observations, revealing the structure of the
molecular outflow in great detail.

\subsubsection{{\rm CO} emission}
\label{co_flux}

Figure~\ref{co_multi} shows the maps of the integrated CO~(\J{1}{0}) emission
in different blueshifted and redshifted velocity intervals. Taking into
account that the systemic velocity, \vlsr, is roughly
$0$~km\,s$^{-1}$ (Patel et al.\ 1995), the interval of
velocities [$\pm3.5$,$\pm9.5$]~km\,s$^{-1}$ corresponds to the low
velocity outflow, [$\pm9.5$,$\pm15.5$]~km\,s$^{-1}$ to the intermediate
velocity outflow, and [$\pm15.5$,$\pm21.5$]~km\,s$^{-1}$ to the high
velocity outflow. Due to the strong self-absorption observed for the
central velocity channels ($-2.8\lesssim v \lesssim 2.3$~km\,s$^{-1}$),
also visible in single-dish observations (Wilking et al.\ 1989;
Codella et al.\ 2001), it was not possible to map and study the cloud
structure. Additionally, extended emission also causes negative contours
due to the fact that the interferometer is filtering out part of the
emission. The CO emission traces at least two molecular outflows in the
region: a stronger outflow along the E--W direction and centered at the
position of BIMA~2, and a collimated, weaker and smaller bipolar outflow
elongated along the N--S direction on both sides of BIMA~1 (see
Fig.~\ref{co_multi}). This latter outflow is only detected at low
velocities, and its presence is also marginally revealed in CO~(\J{2}{1})
single-dish observations of Codella et al.\ (2001). Taking into account
the blueshifted and redshifted emission closer to BIMA~1, the total
extent of this outflow is $\sim 50\arcsec$ ($\sim 0.19$~pc).

The E--W outflow presents a more complex structure and kinematics. At high
velocities the outflow shows a very well defined bipolar morphology, with the
blue lobe toward the southwest, and the red lobe toward the northeast.
However, at lower velocities the blueshifted and redshifted emission are
highly overlapping, and the strongest emission shows a V-shaped morphology that
is probably tracing the bright-rim seen in CO~(\J{2}{1}) single-dish
observations by Codella et al.\ (2001). These authors explain this morphology
as a result of the interaction of the outflow with the ambient cloud (see
$\S~\ref{interaccio}$ for a discussion of the outflow morphology). The
presence of additional outflows in the region could also be a possible
explanation for the complexity of the gas emission. The outflow is highly
collimated, and the collimation remains even at low velocities with respect to
the cloud. The morphology of the outflow at low velocities, with blueshifted
and redshifted gas in both lobes, also suggests that the axis of the outflow
is close to the plane of the sky. In addition to the main structure, there are
also visible at low velocities some blueshifted and redshifted clumps. In
particular, at $\sim 40\arcsec$ northwest from BIMA~2, there is a big
bow-shock shaped clump (see Fig.~\ref{co_multi}) that coincides spatially with
a CS~(\J{2}{1}) clump (see $\S~3.3.2$), suggesting perhaps that there is a
shocked surface at the position of this clump.

\subsubsection{{\rm CS} emission}

The emission of the high-density tracer CS is found to be associated with the
E--W molecular outflow driven by BIMA~2. Serabyn et al.\ (1993) already
detected CS at high velocities through the spectra of different rotational
transitions showing prominent wing emission. However, this is the first time
that this outflow has been mapped with high spatial resolution.
Figure~\ref{cs21_multi} shows the CS~(\J{2}{1}) emission at the same velocity
intervals as shown for the CO~(\J{1}{0}) emission (see previous section). As
can be seen in this figure, the lobes of the outflow are very well defined,
with the redshifted emission NE, and the blueshifted emission toward the SW.
The CS~(\J{2}{1}) emission shows a different morphology at different
velocities, just as in the case of the CO emission. While at high velocities
the outflow is clearly bipolar, the blueshifted and the redshifted emission
start to overlap at lower velocities, and their structures become less
collimated. At low velocities, a redshifted emission clump is also visible in
CS $\sim 40\arcsec$ northwest of BIMA~2, at the same position as the bow-shock
shaped clump seen in CO. In addition, at low and intermediate velocities, a
redshifted clump located $\sim 34\arcsec$ northeast from IRAS~21391+5802 is
detected. Finally, there is also blueshifted emission possibly associated with
BIMA~1. 

The molecular outflow as traced by the CS~(\J{5}{4}) emission shows two well
differentiated and clumpy lobes (see Fig.~\ref{cs_ch3oh_multi}). Both
the blueshifted and redshifted lobes, which have been integrated for a velocity
interval [$\pm 5,\pm12$]~km\,s$^{-1}$, split into several clumps, with the
emission of the northern clump in both lobes being the more prominent and
strong. Hereafter we call these northern blueshifted and redshifted clumps, B
and R, respectively. Additionally, the northeastern clump visible in the
CS~(\J{2}{1}) transition is also detected in CS~(\J{5}{4}) (see
Fig.~\ref{cs_ch3oh_multi}). No outflowing CS~(\J{5}{4}) emission has been found
associated with BIMA~1 nor BIMA~3.

\subsubsection{{\rm CH$_3$OH} emission}

The molecular outflow as traced by CH$_3$OH~(\J{5$_{-1}$}{4$_{-1}$})
resembles in internal structure and extent the emission as traced by 
CS~(\J{5}{4}) (Fig.~\ref{cs_ch3oh_multi}). Both lobes, integrated over [$\pm
5,\pm12$]~km\,s$^{-1}$, are also split into several clumps. The northern
blueshifted and redshifted clumps are also the strongest ones, and their
positions coincide with the positions of the B and R clumps detected in
CS~(\J{5}{4}). The northeastern clump detected in CS~(\J{2}{1}) and
CS~(\J{5}{4}) is not visible in CH$_3$OH. High-velocity gas has not been
detected toward BIMA~1 nor BIMA~3.

\section{Analysis}

\subsection{Dust emission}
\label{dust_analisi}

Using the integrated flux densities at 3.1 and 1.2~mm, we measured the
millimeter spectral index $\alpha$ (where $S_\nu \propto \nu^\alpha$) for the
three sources (see Table~\ref{tmm}). As the emission for BIMA~2 is very
extended, the observations at 1.2~mm, which have a small synthesized beam, are
resolving out part of the emission. Thus, in order to obtain a more accurate
value of the spectral index, it is better to use the flux densities measured
by using the same range of visibilities (5--28~k$\lambda$) at both
wavelengths. By doing this, we measured a spectral index $\alpha<2.3$ for
BIMA~1, $\alpha=3.1\pm0.3$ for BIMA~2, and $\alpha=2.0\pm0.3$ for BIMA~3. For
the sources less resolved (BIMA~1 and BIMA~3) the values obtained do not
differ significantly from the values given in Table~\ref{tmm}. 

Assuming that the dust emission is in the Rayleigh-Jeans regime and is
optically thin, the spectral index $\alpha$ can be related directly to the
power-law index $\beta$ of the dust emissivity $\kappa_\nu$ (where
$\kappa_\nu\propto\nu^{\beta}$), through $\alpha=2+\beta$. For BIMA~2 the
derived value of $\beta$ is $1.1\pm0.3$. Such a value of $\beta$ is consistent
with the values found for Class~0 low- and intermediate-mass objects (Dent,
Matthews, \& Ward-Thompson 1998; Testi \& Sargent 1998; Chini et al.\ 2001).
Saraceno et al.\ (1996) have measured a value $\beta=1.35$ for this source
from ISO-LWS and JCMT data, with an angular resolution of $80\arcsec$ and $\sim
19\arcsec$, respectively. The value measured for BIMA~3 is $\beta=0.0\pm0.3$,
and for BIMA~1 is $\beta<0.3$. Those values are considerably lower than the
value obtained for BIMA~2, and could be indicative of a different evolutionary
status (see \S~\ref{evolution}). Note that Codella et al.\ (2001) have
measured a higher value, $\beta=1.7$, for BIMA~2 and BIMA~3, by combining
together data from both sources obtained with different angular resolution.  

The circumstellar masses for the three sources
were estimated, assuming that the dust emission is optically thin, by using 
\begin{equation} 
M=\frac{c^2}{2\,k\,\nu^2}\,\frac{S_\nu\,d^2}{\kappa_\nu\,T_d},
\end{equation} 
where $S_\nu$ is the flux density, $d$ is the distance,
$\kappa_\nu$ is the dust mass opacity coefficient, and $T_d$ is the dust
temperature. We used the flux densities measured at 3.1~mm, the dust opacity
given by ($\kappa_\nu$/cm$^2$\,g$^{-1}$) $ = 0.1\, (\nu/1200~{\rm GHz})^\beta$
(Hildebrand 1983), with the dust emissivity indices $\beta$ found from the
spectral index (see above), and a dust temperature of 25~K (temperature
estimates for this region range from 20 to 27~K; Serabyn et al.\ 1993; Codella
et al.\ 2001). For BIMA~2, the mass derived was 5.1~$M_\sun$, while for BIMA~1
and BIMA~3 the mass was 0.07~$M_\sun$. The mass estimated for BIMA~2 is
consistent with the value of 8.3~$M_\sun$ derived by Saraceno et al.\ (1996)
through single-dish observations, which do not resolve out extended emission.
However, Codella et al.\ (2001) estimated a mass of 20~$M_\sun$ for BIMA~2,
and 4~$M_\sun$ for BIMA~3. These high values for the mass are due to the
different value of the dust mass opacity coefficient used by them. This shows
clearly how the dust opacity law can affect dramatically the estimate of the
mass.

\subsection{Physical parameters of the CO outflows}

Assuming that the CO emission is in LTE and is optically thin, and following  
the derivation of Scoville et al.\ (1986), we calculated the mass of the gas
associated with the two outflows detected in the region by using
\begin{equation} \label{mass}
\bigg(\frac{M}{M_\sun}\bigg)=2.19\times10^{-5}\,\frac{T_{\rm ex} + 0.93}{{\rm
exp}(-5.59/T_{\rm ex})} \, \bigg(\frac{d}{\rm kpc}\bigg)^2\,
\int{\bigg(\frac{S_\nu}{\rm Jy}\bigg)\,\bigg(\frac{dv}{\rm km\,s^{-1}}\bigg)}, 
\end{equation}  where $T_{\rm ex}$ is the CO excitation temperature, $d$ is the
distance in kiloparsecs, and $S_\nu$ is the line flux density measured in Jy.
The  excitation temperature is 20~K (Codella et al.\ 2001). We assumed an
[H$_2$]/[CO] abundance ratio of 10$^4$, and a mean atomic weight of the gas
$\mu_G=1.36$. In Table~\ref{tco} we give the masses, the CO momentum, the
kinetic energy, and the momentum rate in the outflow. We also report the range
of outflow velocities for the outflows. Note that due to the missing flux, the
absorption by the ambient cloud, and the possibility that CO is optically thick
in portions of the flow, the masses calculated should be considered as lower
limits. In addition, the integration range chosen will miss the part of the
outflow at lower velocities that is not included. We estimated the optical
depth from the observed brightness temperature and by taking $T_{\rm ex}=20$~K,
and we found that our assumption that the emission is optically thin is correct
if beam dilution is unimportant. However, in order to get a better estimate of
the optical depth of the emission, observations of a CO isotope, such as 
$^{13}$CO, would be needed. Note that Codella et al.\ (2001) assume an optical
depth of 10 for the E--W outflow emission at velocities below 20~km\,s$^{-1}$
with respect to the cloud velocity. For such high optical depths, the opacity
correction factor $\frac{\tau}{1-e^{-\tau}}$ is essentially the optical depth,
$\sim 10$, dominating the uncertainty in the measurement of the masses. Thus,
as we cannot rule out the possibility that the gas is clumpy and not optically
thin, the estimated molecular mass may be a lower limit. The inclination angle,
$\theta$, of the outflow with respect to the plane of the sky is also an
important factor of uncertainty when estimating the momentum, the kinetic
energy, and the momentum rate in the outflow. As stated before, the morphology
of the outflow suggests that $\theta$ is small, and thus, we adopted a value
$\theta\simeq15\arcdeg$ (Codella et al.\ [2001] assume a value in the range
10--$20\arcdeg$).

\subsection{Properties of the shocked material}
\label{xoc_material}

We convolved the CS~(\J{5}{4}), CH$_3$OH, and CO maps with a Gaussian to
obtain the same resolution as the CS~(\J{2}{1}) maps, $\sim 7\arcsec$, and
thus, to derive the properties of the shocked material. In
Figure~\ref{cs_spectra} we show the $\sim 7\arcsec$ spectra of the observed
transitions at the CS~(\J{5}{4}) peak intensity position of the B and R
clumps. The spectra at both positions show a remarkable mirror symmetry in
velocity for the CS and the CH$_3$OH transitions. The spectral line profiles
for CS~(\J{2}{1}), which peak near the ambient cloud velocity and then exhibit
a high-velocity blue (B) or red (R) wing, indicate that the accelerated gas
contains a mixture of velocities, and that most of the material is moving at
low outflow velocities (Tafalla \& Bachiller 1995). This mixture of velocities
together with the curved shape of the clumps (see Fig.~\ref{cs_ch3oh_multi})
suggest that the gas has been accelerated by a bow-shock, because in such a
case, the curvature of the bow-shock produces a variety of shock velocities
(e.g.\ Hartigan, Raymond, \& Hartmann 1987). On the other hand, CS~(\J{5}{4})
and CH$_3$OH exhibit a similar profile. The lack of emission of these two
transitions at ambient velocities at the peak position of the clumps B and R
is probably due to a problem of missing flux. Extended emission being filtered
out by the interferometer would cause extended absorption. This is suggested
by the  negative  contours seen clearly for ambient velocities in the channel
maps of CS~(\J{5}{4}) and CH$_3$OH convolved to the same resolution as
CS~(\J{2}{1}) (not presented here), and by the negative fluxes at ambient
velocities (see Fig.~\ref{cs_spectra}). Interestingly, the emission in the
wings of CS~(\J{5}{4}) is roughly constant with velocity, whereas the wings of
CS~(\J{2}{1}) clearly decrease with the outflow velocity. This probably
suggests that the temperature of the high density-gas increases with the
outflow velocity, although it could also be due to an increase of the density
with the outflow velocity. Figure~\ref{21versus54} shows the CS~(\J{2}{1})
spectra and the (\J{5}{4}) to (\J{2}{1}) ratio at the position of the B and R
clumps. This ratio varies from 0.2 for a \vlsr\ of $\pm 3$~\kms\ to 0.6 for a
\vlsr\ of $\pm 14$~\kms. Assuming optically thin emission, this implies a
variation in the excitation temperature from 7 to 12~K, thus the temperature
of the gas traced by CS increases with the outflow velocity. These two
temperatures should be considered as lower limit: first, the optically thin
assumption may not be true, and second, it is possible that the CS~(\J{5}{4})
filling factor is smaller than that of the CS~(\J{2}{1}). Yet, taking into
account that the kinetic temperature measured in the outflow is 20~K (Codella
et al.\ 2001), one would expect much higher kinetic temperatures for the
shocked material. Thus, the emission of the shock is subthermalized, and the
excitation temperature measured is much lower than the real temperature of the
shocked gas.

In order to estimate the CS and \metha\ beam averaged column densities, we
assumed optically thin emission and adopted an excitation temperature of 11~K,
which is the average value of the CS wings. For CO, the adopted excitation
temperature is 20~K (Codella et al.\ 2001). The column densities of CO,
CS, and \metha\ at the peak intensity positions of the B and R clumps are
given in Table~\ref{txoc}. The fractional abundance of CS and \metha\ with
respect to CO at these two positions are shown in Table~\ref{tabun},
together with their typical values for other low- and intermediate-mass
star-forming regions, and for the quiescent and shocked regions in the
low-mass star-forming region L1157. CS and \metha\ show an enhancement of
roughly 2 and 3 orders of magnitude, respectively, with respect to the values
of typical star-forming regions, and similar to the shocked material in the
L1157 molecular outflow.  

\section{Discussion}

\subsection{Interaction of the outflow with the dense material}
\label{interaccio}

Figure~\ref{MapsCSvCO} shows the outflow component of CS~(\J{5}{4})
overlapping with the moderate and high velocity component of the CO~(\J{1}{0})
emission. As can be seen in this figure, the B and R clumps of CS are
located immediately ahead of the blueshifted and redshifted CO emission of the
high outflow velocity component, which traces a well defined bipolar
structure. Further away from the driving source, the molecular outflow is only
detected at moderate or low outflow velocities.

Interestingly, the initially well defined bipolarity of the molecular outflow,
with a well defined axis, appears more chaotic after its passage through these
clumps.  First, west of clump B, the outflow is deflected: from the driving
source to clump B the molecular outflow emission is blueshifted and has a
position angle $\sim 60\arcdeg$, whereas further away from clump B the outflow
is redshifted and its direction changes significantly, to $\sim 110\arcdeg$.
Second, the east lobe emission, between the driving source and clump R, is
mainly redshifted at moderate and high outflow velocities, whereas further
away from clump R the outflow emission is slightly stronger in the blue wing
than in the red wing.

In addition, there is a strong abundance enhancement of CS and \metha\ toward
the positions of clumps B and R (see Table~\ref{tabun}). This is a clear
evidence of strong shocks, where the CS and \metha\ abundances increase from
the evaporation or dust sputtering within the shocks. This has been observed in
other regions, and in particular in L1157 for these two molecules as well as
several others (Mikami et al.\ 1992; Avery \& Chiao 1996; Bachiller \& P\'erez
Guti\'errez 1997; Zhang, Ho, \& Wright 2000). These kinematical and chemical
evidences strongly suggest that the CS clumps are tracing gas entrained within
the surface of interaction between the molecular outflow and the dense ambient
quiescent core. In order to trace and map the dense core material that is
behind the clumps, higher resolution observations of low excitation molecular
lines would be needed. The geometry of the interaction, sketched in
Fig.~\ref{deflection}, suggests the shocked cloudlet model scenario, where an
inverted bow-shock is produced in the windward side of a dense ambient clump
(Schwartz 1978; Davis, Dent, \& Bell Burnell 1990). Here the bow-shock region
is traced by the CS and CH$_3$OH emission (clumps B and R). This scenario is
supported by Figure~\ref{ambient}, where we superposed the integrated intensity
of the CS~(\J{5}{4}) emission at ambient velocities, $[-3,3]~\mbox{km s}^{-1}$,
and the shocked dense gas. As it can be seen in this figure, whereas the
spectra at the positions of the B and R clumps are indicative of shocks, the
spectra at the other two positions, which are located downstream of the clumps
B and R, show a more quiescent shape, with emission peaking at ambient
velocities.

\subsection{Evolution of intermediate-mass stars}
\label{evolution}

The centimeter and millimeter continuum emission around IRAS~21391+5802 has
been resolved into three sources (see Fig.~\ref{centim} and \ref{millim}), with
different morphologies and properties. Although all the sources seem to be
young stellar objects, these differences suggest that they could be in
different evolutionary stages. BIMA~2, the source associated with
IRAS~21391+5802, has the strongest millimeter emission, and it is also the most
massive object in IC~1396N. The source is associated with centimeter continuum
emission, water maser emission (Tofani et al.\ 1995), and it is driving a
powerful molecular outflow. Its millimeter emission, which accounts for most of
the dust emission in the region, exhibits a very extended component, elongated
in a north-south direction, consistent with an envelope surrounding the source.
As seen in $\S~\ref{bima_res}$, the ratio of its intensity peak to its flux
density indicates that there is still plenty of material which can be
incorporated into the central star. The estimated power-law index of the dust
emissivity for BIMA~2, $\beta=1.1\pm0.3$ (see $\S~\ref{dust_analisi}$), is
consistent with the values found for embedded low- and intermediate-mass
objects (Dent et al.\ 1998; Testi \& Sargent 1998; Chini et al.\ 2001). The
circumstellar mass $M\simeq 5.1~M_\sun$, is also consistent with the masses in
the range 3.5--12~$M_\sun$ found around other intermediate-mass protostars
(Fuente et al.\ 2001; Shepherd \& Watson 2002). This mass is also in agreement,
although being slightly higher, with the envelope masses of 1.4--2.3~$M_\sun$
found by Bontemps et al.\ (1996) for low-mass Class~0 sources. Thus, based on
the dust emission results, BIMA~2 seems to have the characteristics of a very
young intermediate-mass object. Note that the morphology and properties of the
dust emission do not differ significantly from the properties of the low-mass
counterpart Class~0 objects.

In order to further compare the properties of BIMA~2 with the low-mass case, 
we also checked whether this object is consistent with some correlations
between source and outflow properties found for low-mass young objects. In
particular, we checked for the correlation between the circumstellar envelope
mass and the momentum rate in the CO outflow given by Bontemps et al.\ (1996).
To do this, first we corrected the momentum rate for projection and optical
depth effects using the same correction factor as Bontemps et al.\ (1996), by
taking $\dot P\simeq 10\times \dot P_{\rm obs}$, where $\dot P_{\rm obs}$ is
the observed momentum rate. We found that BIMA~2 agrees well with this
correlation, and, what is more, its position in the diagram is consistent with
the source being a Class~0 object, which confirms the classification of BIMA~2
as a very YSO. These authors have also found a correlation between the
normalized outflow momentum rate, $\dot P\,c/L_{\rm bol}$, and the normalized
envelope mass, $M/L_{\rm bol}^{0.6}$. For IRAS~21391+5802, there are some
bolometric luminosity estimates, $L_{\rm bol}\simeq235$--440~$L_\sun$ (Saraceno
et al.\ 1996; Sugitani et al.\ 2000). However, these estimates include the
whole region around the IRAS coordinates. Thus, in order to derive a bolometric
luminosity for BIMA~2, we used the relationship between the momentum rate and
the bolometric luminosity given by Cabrit \& Bertout (1992), and we inferred a
luminosity of $L_{\rm bol}\simeq 150~L_\sun$. By using this luminosity estimate
we found that the source also fits very well the correlation of normalized
momentum and envelope mass as found by Bontemps et al.\ (1996), and BIMA~2 also
lies in the Class~0 region of this diagram. 

Another correlation fitted well by the source is the one  between the radio
continuum luminosity at centimeter wavelengths and the momentum rate of the
outflow not corrected for inclination obtained by Anglada (1996). This
correlation is in agreement with the predictions of a simple model of shock
ionization in a plane-parallel geometry (Curiel, Cant\'o, \& Rodr\'{\i}guez
1987; Curiel et al.\ 1989), which would be able to produce the required
ionization in thermal radio jets. Regarding the morphology of the outflow, at
low outflow velocities it is very complex and chaotic (see
Fig.~\ref{co_multi}), as also noticed in other intermediate-mass star-forming
regions (e.g.\ NGC~7129: Fuente et al.\ 2001). However, unlike the 
NGC~7129~FIRS1 outflow, the collimation of the BIMA~2 outflow is very high,
even at low velocities, similar to the low-mass protostellar flow in HH~211
(Gueth \& Guilloteau 1999). The complexity of the molecular outflows driven by
intermediate-mass protostars is likely a result of the more complex
protostellar environment itself. Taking into account that the outflows are
usually more energetic for higher mass objects, that the dust emission is
often resolved into more than one object, and that these sources are embedded
in larger amounts of material, it is to be expected that the interactions
between the high-velocity gas and the circumstellar material will be more
active and more complex, disrupting and pushing more material. As seen in
$\S~\ref{xoc_material}$, the chemistry of the shocked material in the outflow,
which shows an enhancement in the abundance of CS and CH$_3$OH, is different
from that of the quiescent material in the cloud. Interestingly, the
abundances of these species are quite similar for all shocked regions,
regardless of whether they are related to intermediate- or low-mass
star-forming regions such as L1157 (Bachiller \& P\'erez Guti\'errez 1997).  

BIMA~3 is the other source for which the millimeter continuum emission has
been resolved, although only at 1.2~mm. Its emission is more compact than the
emission around BIMA~2, and has a deconvolved size of $\sim 1500$~AU, which is
still consistent with envelope sizes. The source is associated with centimeter
continuum emission and with water maser emission (Tofani et al.\ 1995).  The
more compact appearance of the dust emission, together with the fact that the
source does not seem to be associated with any molecular outflow suggest that
BIMA~3 might be a less massive and/or more evolved object than BIMA~2. On one
hand, the circumstellar mass obtained for this source is consistent with the
values found for low-mass objects (Bontemps et al.\ 1996; Motte \& Andr\'e
2001). On the other hand, the evolved stage seems to be confirmed by the
spectral index, $\alpha=2.0\pm0.3$, obtained at millimeter wavelengths for
this source. In order to obtain this value, the millimeter emission would have
to be either optically thick, or optically thin with a dust emissivity index
$\beta=0.0\pm0.3$. Although the deconvolved size of the emission depends on
the angular resolution (see $\S~\ref{bima_res}$), the size derived, even the
one derived with the highest resolution, is too large to interpret this
emission as an optically thick disk-like structure, which would be
gravitationally unstable. Typical circumstellar dusty disk radii are on the
order of 100~AU for T~Tauri stars (e.g.\ Dutrey et al.\ 1996), and in the
range 100--300~AU for Herbig~Ae stars (Mannings \& Sargent 1997, 2000; Natta
et al.\ 2000). In addition, such a structure would be excessively massive.
Perhaps a more plausible explanation is that the emission is optically thin
and $\beta=0.0\pm0.3$. A smaller value of $\beta$ around compact objects has
been interpreted as caused by larger grains, indicating grain growth in dense
regions (Mannings \& Emerson 1994); grain shape evolution or chemical
evolution are also possible causes (Ossenkopf \& Henning 1994; Pollack et al.\
1994). 

The third source in the region, BIMA~1, is also associated with centimeter
continuum emission, and it is probably the driving source of the collimated
N--S bipolar outflow discovered in the region (see $\S~\ref{co_flux}$). These
observational facts indicate that BIMA~1 is also a YSO, although less massive
than BIMA~2. However, due to the fact that its millimeter continuum emission
has not been resolved, it is not possible to say too much about its
evolutionary stage. The spectral index obtained for this source is
$\alpha<2.3$, which in the case that the millimeter continuum emission is
optically thin would correspond to a low value of $\beta<0.3$, and might be
interpreted as grain growth. However, as the deconvolved size of BIMA~1 is
$<1100$~AU at 1.2~mm, we cannot rule out the possibility that what we are
seeing is an optically thick circumstellar disk (e.g.\ L1641-N: Chen, Zhao, \&
Ohashi 1995). Assuming $\tau=1$, $\kappa_\nu=0.01$~cm$^2$\,g$^{-1}$ at 3.1~mm,
and $\beta=1$ (Beckwith et al.\ 1990), we obtain a disk mass $>0.35\,M_\sun,$
for a disk radius of 100~AU. In case the emission is optically thin, the value
derived for the circumstellar mass is $M=0.07~M_\sun$, a value consistent with
a low-mass object.

\section{Conclusions}

We studied with the VLA and the BIMA arrays the dust and the gas at
centimeter and millimeter wavelengths surrounding IRAS~21391+5802, an
intermediate-mass source embedded in the core of IC~1396N.

The continuum emission at centimeter and millimeter wavelengths has been
resolved into three sources, BIMA~1, BIMA~2, BIMA~3, separated $\sim 15\arcsec$
of each other. The strongest source at millimeter wavelengths is BIMA~2, which
is most likely the object associated with IRAS~21391+5802, and its dust
emission shows a very extended envelope elongated in the N--S and the NE--SW
direction at 3.2~mm and 1.3~mm, respectively. The power-law index of the dust
emissivity measured from 3.1 to 1.2~mm is $\beta=1.1\pm0.3$ for BIMA~2, and its
circumstellar mass is $5.1~M_\sun$. The dust emission around BIMA~1 has not
been resolved, and the emission around BIMA~3 has only been resolved at 1.2~mm.
The dust emissivity index is $\beta<0.3$ for BIMA~1, and $\beta=0.0\pm0.3$ for
BIMA~3, and the circumstellar mass is $0.07~M_\sun$ for both objects.

The CO emission traces at least two molecular outflows in the region: a
stronger E--W outflow centered at the position of BIMA~2, and a collimated,
weaker, and smaller outflow elongated along the N--S direction on both sides of
BIMA~1. The morphology and kinematics of the E--W outflow are very complex:
while at high outflow velocities its morphology is clearly bipolar, at low
outflow velocities the blueshifted and redshifted emission are highly
overlapping, and the emission is deflected showing a V-shaped morphology. Unlike
other outflows driven by intermediate-mass objects, the E--W outflow is highly
collimated, and the collimation remains even at low outflow velocities.

The E--W outflow as traced by CS and CH$_3$OH shows two well differentiated
and clumpy lobes, with two prominent curved northern blueshifted and
redshifted clumps. CS and the CH$_3$OH are strongly enhanced toward the
position of these clumps with respect to typical quiescent material abundances
in other star-forming regions. The CS~(\J{5}{4}) to (\J{2}{1}) ratio at both
positions increases from 0.2 for a \vlsr\ of $\pm 3$\kms\ to 0.6 for a \vlsr\
of $\pm 14$\kms, indicating that the excitation temperature of the gas traced
by CS increases with the outflow velocity. All these morphological,
kinematical and chemical evidences suggest that the clumps are tracing gas
entrained within the surface of interaction between the molecular outflow and
the dense ambient quiescent core. They also suggest that the morphology of the
molecular outflow, which is being deflected toward the north at low outflow
velocities, is a result of this interaction.

All the objects in IC~1396N seem to be young stellar objects, but in
different evolutionary stages. The youngest one is BIMA~2: its circumstellar
mass together with the dust emissivity index, and the fact that the source is
driving a molecular outflow are consistent with the source being an embedded
intermediate-mass protostar. In addition, the source also fits very well
correlations between source and outflow properties for low-mass Class~0
objects given by Bontemps et al.\ (1996). By using the relationship between
the momentum rate and the bolometric luminosity (Cabrit \& Bertout 1992), we
inferred a bolometric luminosity of $\sim 150~L_\sun$ for this source. For
BIMA~1 and BIMA~3, the small dust emissivity index measured is suggestive of
grain growth in dense regions (e.g.\ Mannings \& Emerson 1994). This fact
together with the more compact appearance of their dust emission suggest that
they are more evolved low-mass objects.

As compared with the low-mass counterparts, the morphology and properties of
the dust emission around the intermediate-mass protostar BIMA~2 do not differ
significantly. In addition, the properties of the outflow driven by BIMA~2 are
also consistent with those of the outflows driven by low-mass young objects
(Bontemps et al.\ 1996). The complexity of the molecular outflow, as compared
with the low-mass case, is likely a result of the more complex protostellar 
environment itself. Interestingly, the properties of the shocked regions are
quite similar to the ones related to low-mass star-forming regions. The
correlation between its flux density at centimeter wavelengths and the momentum
rate in the CO outflow agrees with the one found by Anglada (1996) for thermal
radio jets, which can be understood under the shock ionization model of Curiel
et al.\ (1987, 1989). 

\acknowledgments

We would like to thank Qizhou Zhang for his useful help and comments on
interferometric spectral line observations. We also thank N\'uria Calvet,
Nimesh Patel and the anonymous referee for their valuable comments. We
acknowledge the anonymous referee for useful comments. M.\ T.\ B.\ is supported
by a SAO predoctoral fellowship. J.\ M.\ G.\ acknowledges support by NSF grant
AST-99-81363 and by RED-2000 from the Generalitat de Catalunya. A.\ P.\
acknowledges support by FI-2001 grant from DURSI of the Generalitat de
Catalunya. R.\ E., J.\ M.\ G., and A.\ P.\ are partially supported by MCYT
grant PB98-0670 (Spain).

{}

\clearpage

\subsection*{Figure captions}

\figcaption{VLA natural-weight map of the 3.6~cm emission (solid
contours) overlaid with an optical image from the Digitized Sky Survey 2
(grey levels). The resulting synthesized beam, shown in the upper right
corner, is $17\farcs9\times8\farcs8$ with $\rm P.A.=-80\arcdeg$.  The
contour levels are $-10$, $-7$, -4, 4, 7, 10, 20, 40, 70, 110, and 130
times the rms of the map, $20~\mu \mbox{Jy beam}^{-1}$.}

\figcaption{VLA uniform-weight map at 3.6~cm. The synthesized
beam, $12\farcs6\times5\farcs9$ with $\rm P.A.=-82\arcdeg$, is shown in
the upper right corner. The contour levels are $-3$, 3, 6, 9, and 12
times the rms of the map, $32~\mu$Jy beam$^{-1}$. The error ellipse of
IRAS 21391+5802 is indicated. Plus signs correspond to near infrared
sources of Wilking et al.\ (1993).}

\figcaption{BIMA map of the 3.1~mm (top panel), and the 1.2~mm (bottom panel)
continuum emission toward the core of IC~1396N. The synthesized beam is
$8\farcs0\times6\farcs7$ with $\rm P.A.=-14\arcdeg$ for the 3.1~mm map, and
$1\farcs5\times1\farcs3$ with $\rm P.A.=7\arcdeg$ for the 1.2~mm map. The
contour levels are $-3$, 3, 6, 9, 12, 15, 20, 25, 30, and 40 times
$1.4~\mbox{mJy beam}^{-1}$ for the 3.1~mm map, and $2.0~\mbox{mJy beam}^{-1}$
for the 1.2~mm map. The error ellipse of IRAS 21391+5802 is indicated. Crosses
show the positions of the 3.6~cm sources (this paper).}

\figcaption{CO~(\J{1}{0}) emission integrated in different velocity intervals,
$[\pm3.5,\pm9.5]~\mbox{km s}^{-1}$ (top panel), $[\pm9.5,\pm15.5]~\mbox{km
s}^{-1}$ (middle panel), and $[\pm15.5,\pm21.5]~\mbox{km s}^{-1}$ (bottom
panel). The thick contours depict the blueshifted integrated emission with
respect to the systemic velocity, and the thin contours the redshifted
emission. The synthesized beam, $4\farcs3\times3\farcs8$ with  $\rm
P.A.=11\arcdeg$, is shown in the lower left corner. The black dots show the
positions of the 3.1~mm sources (this paper). The contours are 3, 5, 10, and
15 times $(a)$ $0.6~\mbox{Jy beam}^{-1}$ $\mbox{km s}^{-1}$, $(b)$
$0.3~\mbox{Jy beam}^{-1}$ $\mbox{km s}^{-1}$, and $(c)$ $0.3~\mbox{Jy
beam}^{-1}$ $\mbox{km s}^{-1}$.}

\figcaption{CS~(\J{2}{1}) emission integrated in different velocity intervals,
$[\pm3.5,\pm9.5]~\mbox{km s}^{-1}$ (top panel), $[\pm9.5,\pm15.5]~\mbox{km
s}^{-1}$ (middle panel), and $[\pm15.5,\pm21.5]~\mbox{km s}^{-1}$
(bottom panel). The thick contours depict the blueshifted integrated emission
with respect to the systemic velocity, and the thin contours the redshifted
emission. The synthesized beam, $7\farcs0\times6\farcs3$ with  $\rm
P.A.=-21\arcdeg$, is shown in the lower left corner. The symbols are as in
Fig.\ 4. The contours are 3, 5, 10, and 15 times $0.35~\mbox{Jy beam}^{-1}$
$\mbox{km s}^{-1}$.}

\figcaption{Integrated intensity of the  CS~(\J{2}{1}) emission (top panel) in
a velocity interval $[\pm5,\pm35]~\mbox{km s}^{-1}$, and a synthesized beam of
$7\farcs0\times6\farcs3$ with P.A.=$-21\degr$, CS~(\J{5}{4}) emission (middle
panel) in a velocity interval $[\pm5,\pm12]~\mbox{km s}^{-1}$, and a
synthesized beam of $2\farcs1\times1\farcs8$ with P.A.=$-3\degr$, and
CH$_3$OH~(\J{5}{4}) emission (bottom panel) in a velocity interval
$[\pm5,\pm12]~\mbox{km s}^{-1}$, and a synthesized beam of
$2\farcs1\times1\farcs8$ with P.A.=$-1\degr$. The synthesized beam is shown in
the lower left corner. The thick contours depict the blueshifted integrated
emission with respect to the systemic velocity, and the thin contours the
redshifted emission. The symbols are as in Fig.\ 4. The contours are 3, 5, 10,
15, and 20 times $(a)$ $0.55~\mbox{Jy beam}^{-1}$ $\mbox{km s}^{-1}$, $(b)$
$0.26~\mbox{Jy beam}^{-1}$ $\mbox{km s}^{-1}$, and  $(c)$ $0.26~\mbox{Jy
beam}^{-1}$ $\mbox{km s}^{-1}$.}

\figcaption{Spectra of the molecules observed taken at the CS~(\J{5}{4}) peak
intensity position of the clumps B (left panels) and R (right panels).}

\figcaption{CS~(\J{2}{1}) spectrum (solid line) versus the (\J{5}{4}) to
(\J{2}{1}) ratio (dashed line) at the position of the B (left panel) and R
(right panel) clumps. The enhancement of the lines ratio with outflow velocity
is not seen away from these strongly shocked regions.}

\figcaption{Overlap of the shocked dense gas traced by the CS~(\J{5}{4}) line
(grey scale) and the molecular outflow traced by the CO~(\J{1}{0}) line (solid
and dashed contours for the blueshifted and redshifted emission), at high (top
panel) and intermediate (bottom panel) outflow velocities. CO and CS maps are
the same as those on Fig.~\ref{co_multi} and \ref{cs_ch3oh_multi}
respectively.}

\figcaption{Sketch showing the geometry of the interaction of the CO outflow
with the dense material, which is traced by CS and CH$_3$OH.}

\figcaption{Superposition of the integrated intensity of the CS~(\J{5}{4})
emission over the velocity interval $[-3,3]~\mbox{km s}^{-1}$ (contours), and
the shocked dense gas (grey scale). The contours of the ambient gas emission
are 3, 6, 9, 12, 16, 20, 25, and 30 times 0.12~$\mbox{Jy beam}^{-1}$. The black
dots show the positions of the 3.1~mm sources (this paper). The crosses
indicate the positions where the spectra (top panel) have been taken.}

\clearpage

\begin{figure}[hbt]
\figurenum{1}
\plotone{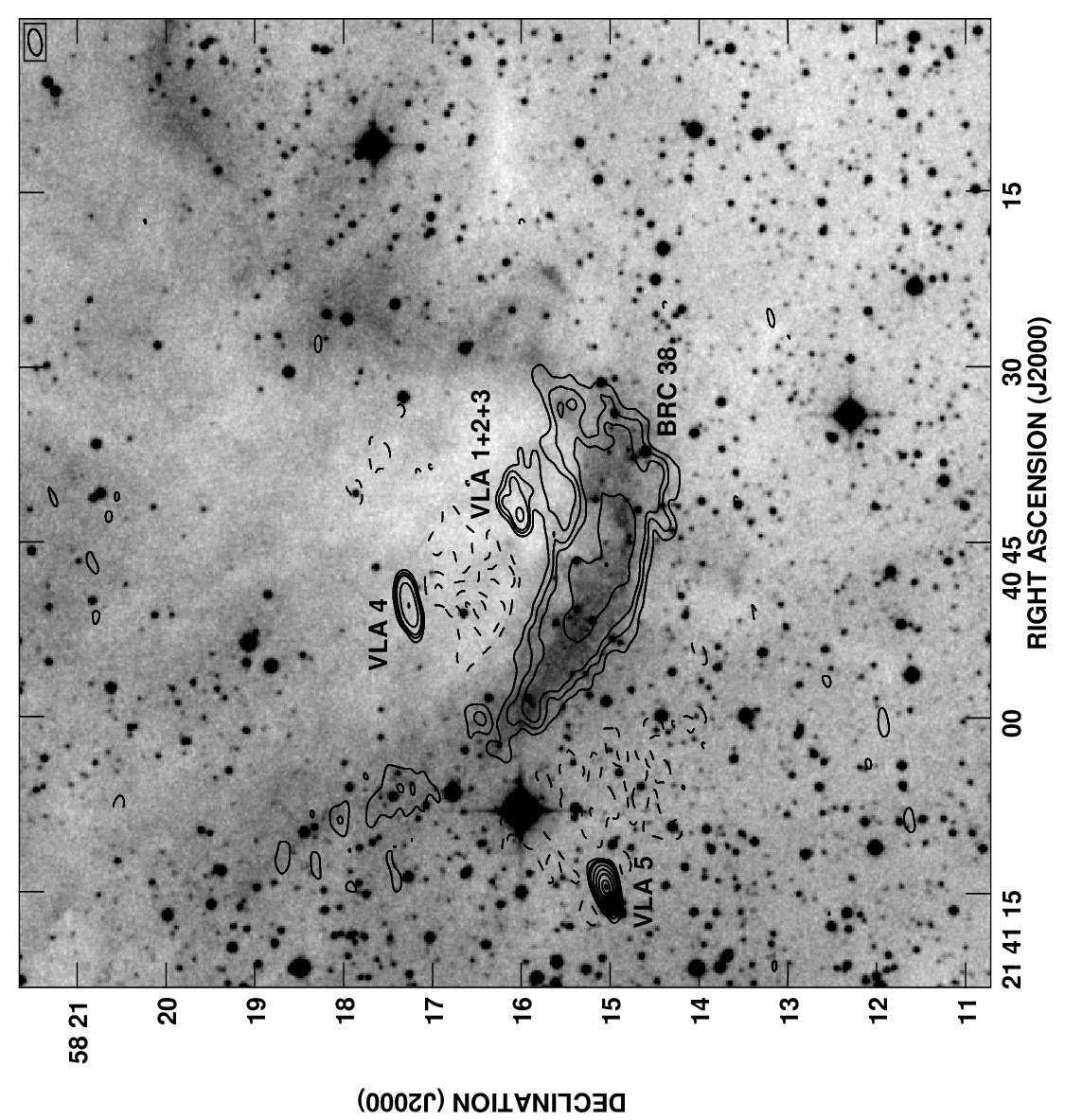}
\caption{}   
\end{figure}
\clearpage
\begin{figure}[hbt]
\figurenum{2}
\label{centim}
\plotone{f2.ps}
\caption{}
\end{figure}
\clearpage
\begin{figure}[hbt]
\figurenum{3}
\label{millim}
\epsscale{1.5}
\plottwo{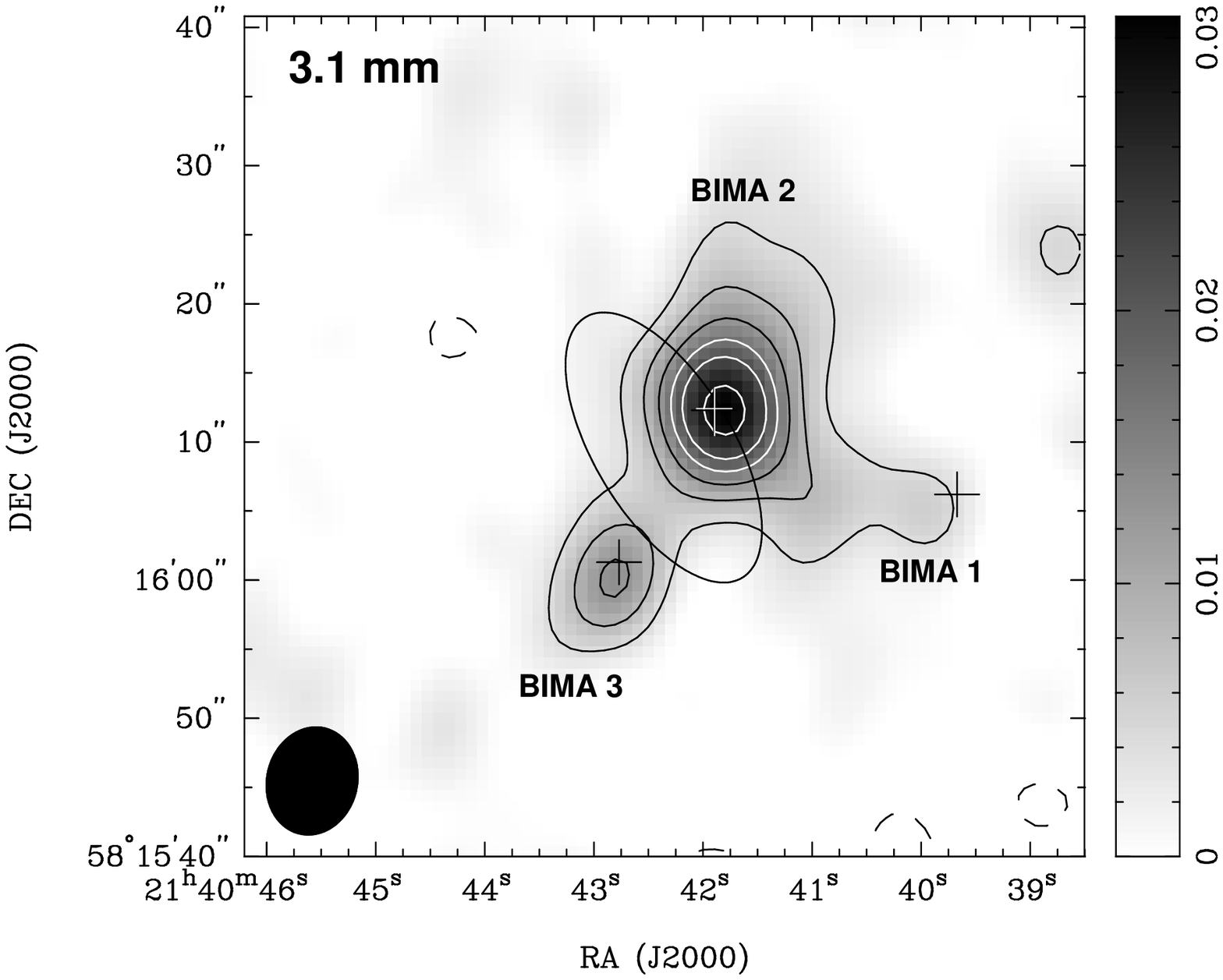}{f3b.ps}
\caption{}
\end{figure}
\clearpage
\begin{figure}[hbt]
\figurenum{4}
\label{co_multi}
\epsscale{0.7}
\plotone{f4.ps}
\caption{}
\end{figure}
\clearpage
\begin{figure}[hbt]
\figurenum{5}
\label{cs21_multi}
\epsscale{0.75}
\plotone{f5.ps}
\caption{}
\end{figure}
\clearpage
\begin{figure}[hbt]
\figurenum{6}
\label{cs_ch3oh_multi} 
\epsscale{0.75}
\plotone{f6.ps}
\caption{}
\end{figure}
\clearpage
\begin{figure}
\figurenum{7}
\label{cs_spectra}
\epsscale{0.9}
\plotone{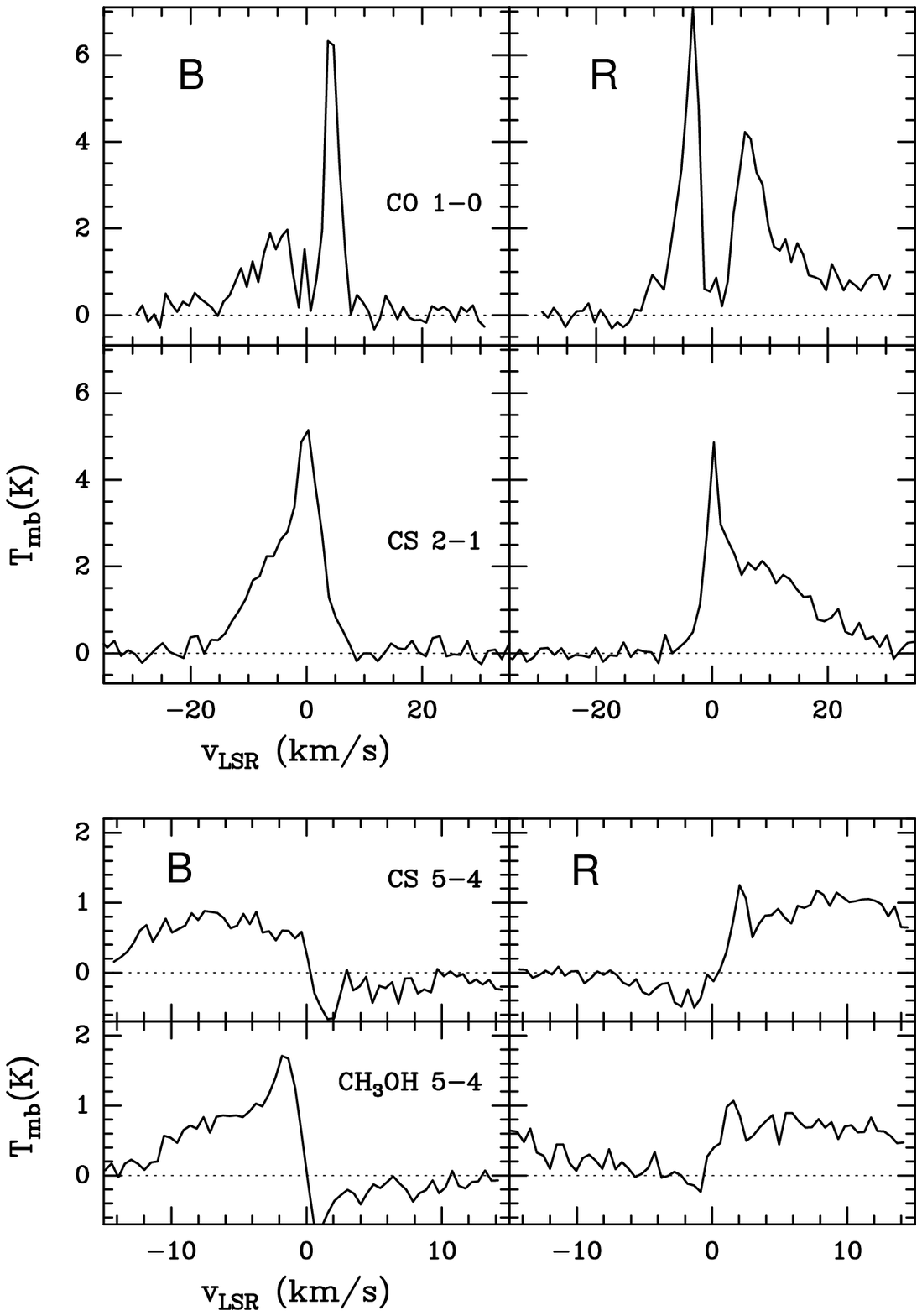}
\caption{}
\end{figure}
\clearpage
\begin{figure}
\figurenum{8}
\label{21versus54}
\epsscale{0.9}
\plotone{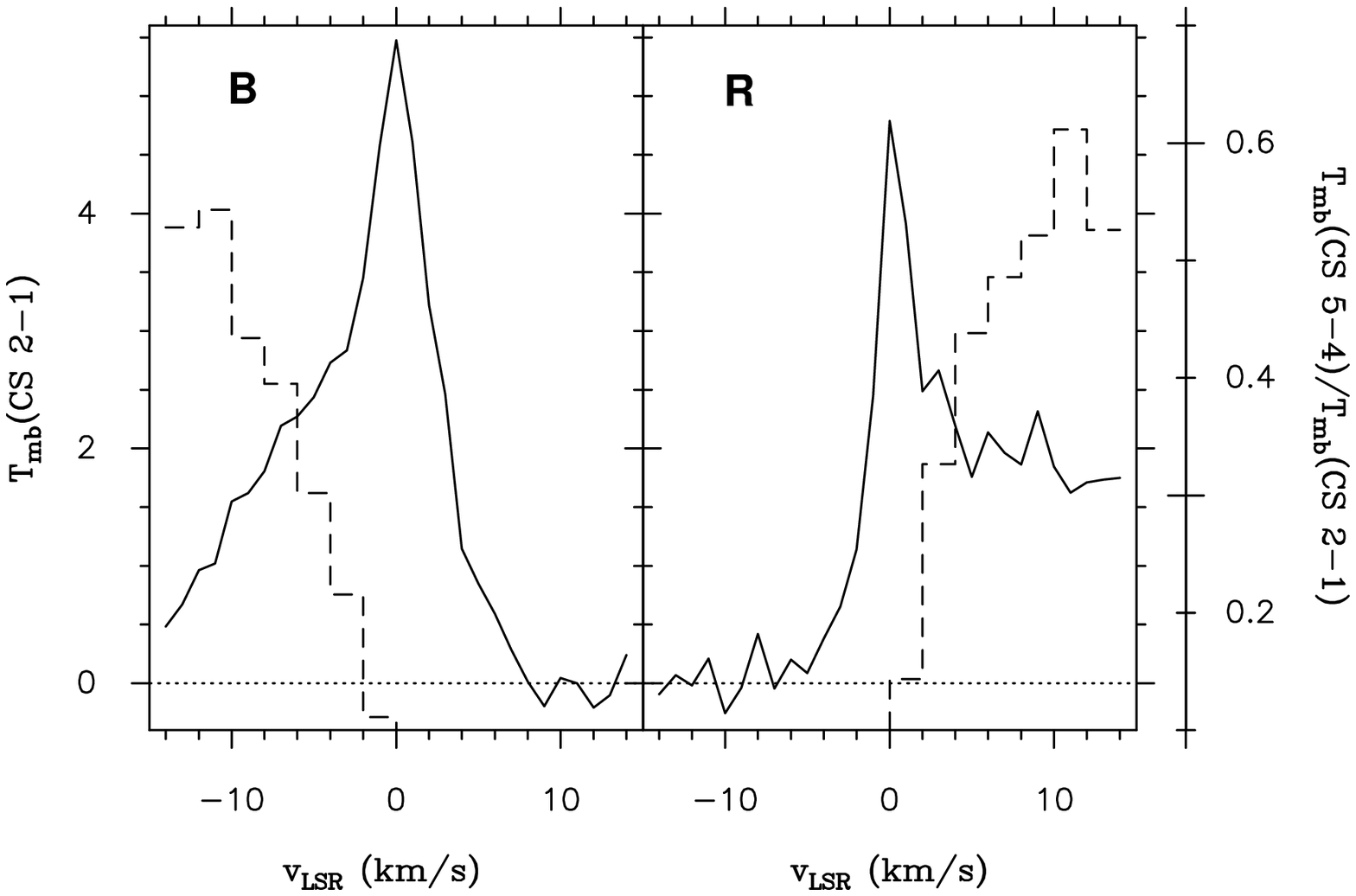}
\caption{}
\end{figure}
\clearpage
\begin{figure}[hbt]
\figurenum{9}
\label{MapsCSvCO}
\epsscale{0.9}
\plotone{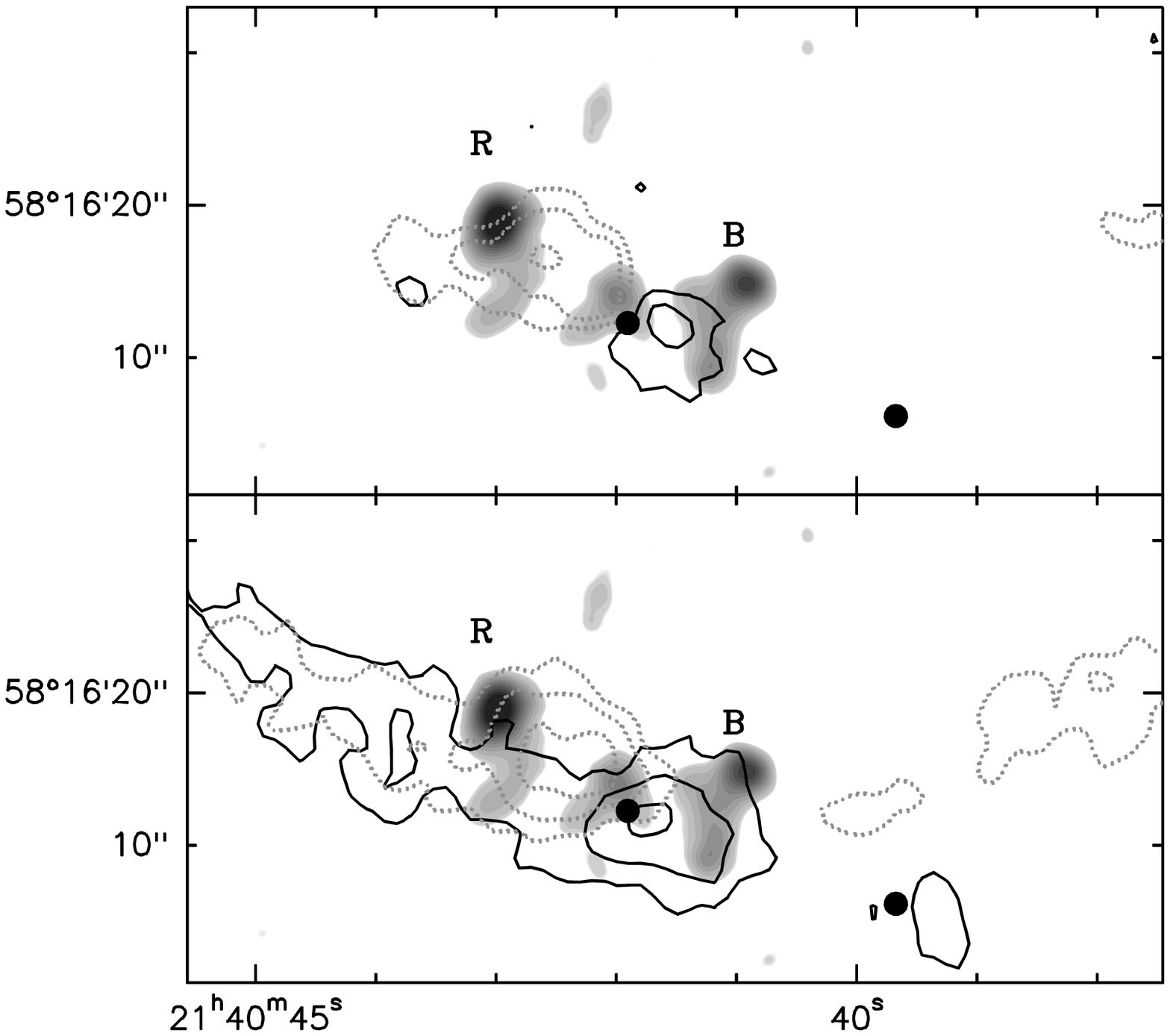}
\caption{}
\end{figure}
\clearpage
\begin{figure}[hbt]
\figurenum{10}
\label{deflection}
\epsscale{0.9}
\plotone{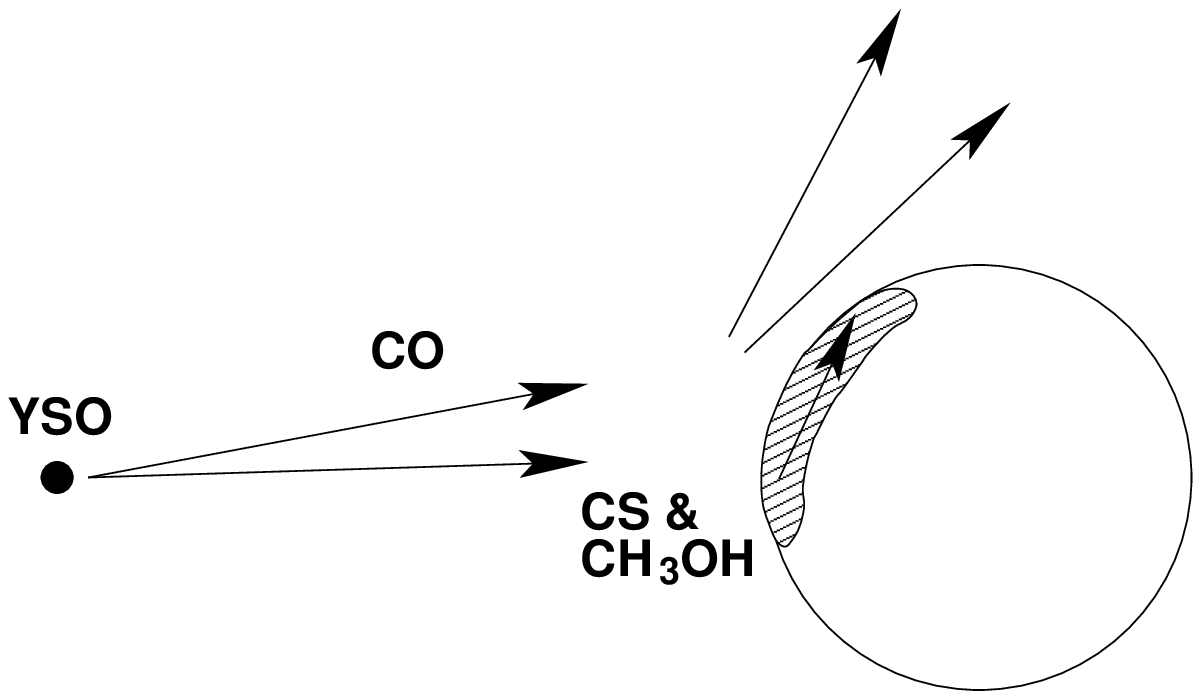}
\caption{}
\end{figure}
\clearpage
\begin{figure}[hbt]
\figurenum{11}
\label{ambient}
\epsscale{0.9}
\plotone{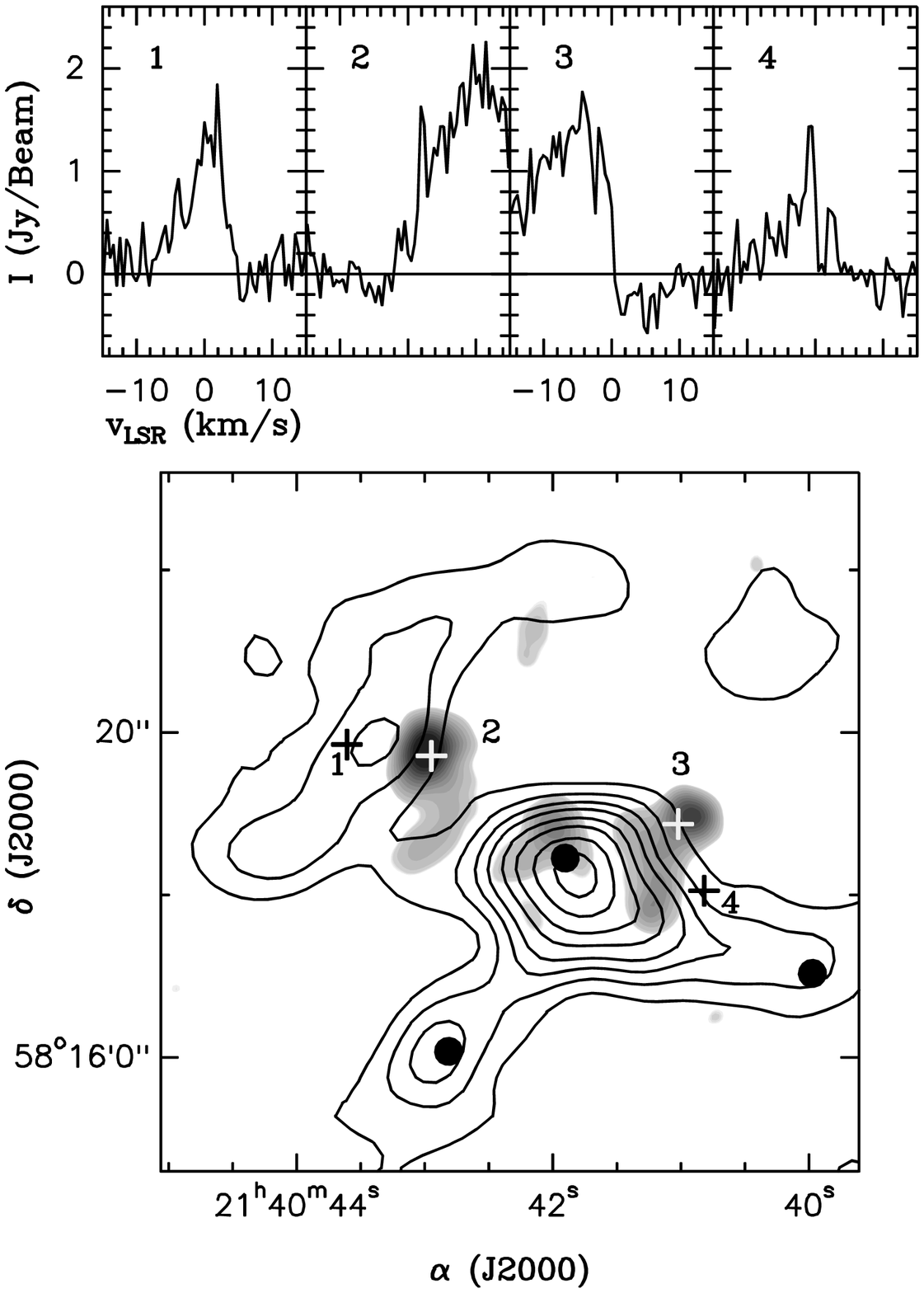}
\caption{}
\end{figure}

\clearpage

\begin{deluxetable}{lclcccrrcc}
\tabletypesize{\scriptsize}
\tablewidth{0pt}
\tablecaption{Parameters of the BIMA observations
\label{tbima}}
\tablehead{
&&&&\colhead{BL Lac\tablenotemark{a}}
&&&&\colhead{Spectral} \\  
&&&\colhead{$\nu$}&
\colhead{Flux}&
\multicolumn{2}{c}{Synthesized Beam}&
\colhead{Bandwidth}&
\colhead{resolution}&
\colhead{T$_{\rm sys}$\tablenotemark{b}}
\\
\cline{6-7}
\colhead{Epoch}&
\colhead{Config.}&
\colhead{Observation}&
\colhead{(GHz)}&
\colhead{(Jy)}&
\colhead{HPBW}&
\colhead{P.A.}&
\colhead{(MHz)}&
\colhead{(km\,s$^{-1}$)}&
\colhead{(K)}
}
\startdata
2000 May 12 &C &continuum &96.25\phn\phn 
&3.8&$8\farcs0\times6\farcs7$
&$-14\degr$ &1200 &&160--500\phn\\
&& CS 2--1 &97.9810 &&$7\farcs0\times6\farcs3$ 
&$-21\degr$ &25 &0.30\\
2000 Mar 04 &B &continuum &112.2\phn\phn\phn\phn 
&4.8 &$2\farcs7\times2\farcs0$
&$40\degr$ &725 &&400--980\phn\\
&& CO 1--0 &115.2712\phn &&$2\farcs7\times2\farcs1$
&$28\degr$ &25 &0.51\\
2000 Apr 05 &C &continuum &112.2\phn\phn\phn\phn 
&3.1&$6\farcs8\times6\farcs3$
&$-17\degr$ &725 &&350--930\phn\\
&& CO 1--0 &115.2712\phn &&$6\farcs6\times6\farcs0$
&$-17\degr$ &25 &0.51\\
2000 Nov 17 &C &continuum &243.3\phn\phn\phn\phn 
&1.8&$3\farcs9\times3\farcs1$
&$-47\degr$ &1200 &&500--1200\\
&& CH$_3$OH 5$_{-1}$--4$_{-1}$ &241.7672\phn &&$3\farcs7\times3\farcs1$
&$-51\degr$ &25 &0.12\\
&& CS 5--4 &244.9356\phn &&$3\farcs7\times3\farcs0$
&$-50\degr$ &25 &0.12\\
2001 Jan 28 &B &continuum &243.3\phn\phn\phn\phn 
&1.9&$1\farcs0\times0\farcs9$
&$11\degr$ &1200 &&270--550\phn\\
&& CH$_3$OH 5$_{-1}$--4$_{-1}$ &241.7672\phn &&$1\farcs1\times0\farcs9$
&$12\degr$ &25 &0.12\\
&& CS 5--4 &244.9356\phn &&$1\farcs1\times0\farcs9$
&$12\degr$ &25 &0.12\\   
\enddata

\tablenotetext{a}{Phase calibrator.}
\tablenotetext{b}{Range of system temperatures during the observations.}
\end{deluxetable}

\clearpage

\begin{deluxetable}{lcccccc}
\tabletypesize{\scriptsize}
\tablewidth{0pt}
\tablecaption{Parameters of the Radio Continuum Sources Detected in the
IC~1396N Region
\label{tvla}}
\tablehead{
&&&\colhead{Integrated Flux}
&\colhead{Integrated Flux}
\\
&\multicolumn{2}{c}{Position\tablenotemark{a}}
&\colhead{Density at 3.6 cm\tablenotemark{b}}
&\colhead{Density at 21 cm\tablenotemark{c}}  
&\colhead{Spectral}\\
\cline{2-3}
\colhead{Source}
&\colhead{$\alpha (\rm J2000)$}
&\colhead{$\delta (\rm J2000)$}
&\colhead{(mJy)} 
&\colhead{(mJy)}
&\colhead{Index\tablenotemark{d}}
&\colhead{Identification}
}  
\startdata
\,(1)
&$21^{\rm h}40^{\rm m}39\fs67$
&$+58\arcdeg16\arcmin06\farcs2$
&$0.21\pm0.03$  
&$<2.0$\tablenotemark{e}
&$>-1.3$  &BIMA 1
\\
\,(2)
&$21^{\rm h}40^{\rm m}41\fs90$
&$+58\arcdeg16\arcmin12\farcs3$
&$0.27\pm0.03$
&$<2.0$\tablenotemark{e}
&$>-1.2$
&IRAS 21391+5802, BIMA 2
\\
\,(3)
&$21^{\rm h}40^{\rm m}42\fs77$
&$+58\arcdeg16\arcmin01\farcs3$
&$0.43\pm0.03$
&$<2.0$\tablenotemark{e}
&$>-0.9$
&BIMA 3
\\
\,(4)
&$21^{\rm h}40^{\rm m}50\fs42$
&$+58\arcdeg17\arcmin16\farcs4$
&$1.84\pm0.03$
&$\phn8.3\pm0.5$
&$-0.83\pm0.04$
\\
\,(5)
&$21^{\rm h}41^{\rm m}14\fs50$
&$+58\arcdeg15\arcmin03\farcs0$
&$32\pm7\phn$
&$24.8\pm0.5$
&\phs$0.14\pm0.12$
\\
\,Extended
\\
\,emission
&$21^{\rm h}40^{\rm m}45\fs09$
&$+58\arcdeg15\arcmin19\farcs4$
&$26\pm2$\tablenotemark{f}\phn 
&$27.6\pm2.0$\tablenotemark{f}
&$-0.03\pm0.06$
&BRC 38
\enddata
\tablenotetext{a}{Typical uncertainty in position is $0\farcs2$.}
 
\tablenotetext{b}{Corrected for primary beam response.}
   
\tablenotetext{c}{From the NRAO VLA Sky Survey (Condon et al.\ 1998).}

\tablenotetext{d}{Spectral index between 21 and 3.6~cm.}

\tablenotetext{e}{Upper limit for non-detected sources is 4$\sigma$.}

\tablenotetext{f}{Flux density integrated inside a box of 
$4\farcm7\times2\farcm1$ with center
given in the position columns. Errors are calculated as $\sigma\sqrt{A}$,
where $A$ is the box to beam areas ratio (see Beltr\'an et al. 2001).}

\end{deluxetable}

\clearpage

\begin{deluxetable}{lrrcccccrcc}
\tabletypesize{\scriptsize}
\tablewidth{0pt}
\tablecaption{Millimeter Flux Densities, Sizes, Spectral Indices and
Masses 
\label{tmm}}
\tablehead{
&
\colhead{$\alpha$ (J2000)\tablenotemark{a}}&
\colhead{$\delta$ (J2000)\tablenotemark{a}}&
\colhead{$\lambda$}&
\colhead{$I_{\nu}^{\rm peak}$\tablenotemark{b}}&
\colhead{$S_{\nu}$\tablenotemark{c}}&
\multicolumn{3}{c}{Source size\tablenotemark{d}}&
&\colhead{Mass\tablenotemark{f}}
\\
\cline{7-9}
\colhead{Source}&
\colhead{($^{\rm h}$ $^{\rm m}$ $^{\rm s}$)}&
\colhead{($\arcdeg$ $'$ $''$)}&
\colhead{(mm)}&
\colhead{(mJy beam$^{-1}$)}&
\colhead{(mJy)}&
\colhead{($\arcsec$)}&
\colhead{(AU)}&
\colhead{P.A.}&
\colhead{$\alpha$\tablenotemark{e}}&
\colhead{($M_\sun$)}
}
\startdata
BIMA~1 &21 40 39.97 &58 16 05.2 &3.1
&\phn$6.3\pm1.3$ &$6.3\pm1.3$
&$<7.3$\tablenotemark{g} 
&$<5500$\tablenotemark{g} &\nodata\tablenotemark{g} 
&$<2.4$ & 0.07 \\
&&&1.2 &$<13.1$\tablenotemark{h} &$<13.1$\tablenotemark{h}
&$<1.4$\tablenotemark{g}
&$<1100$\tablenotemark{g} &\nodata\tablenotemark{g} \\
BIMA~2 &21 40 41.79 &58 16 12.4 &3.1 &$31.0\pm6.2$
&$61.7\pm12.3$
&$7.3\times4.9$ &$5500\times3700$ &$10\arcdeg$ 
&$2.5\pm0.3$ &5.1\\
&&&1.2 &$\phn94.1\pm18.8$\tablenotemark{i}
&$650\pm130$ &$2.6\times2.1$ 
&$2000\times1600$ &$-21\arcdeg$ \\
BIMA~3 &21 40 42.81 &58 16 00.4 &3.1 &$13.8\pm2.8$
&$13.8\pm2.8$\phn &$<7.3$\tablenotemark{g}
&$<5500$\tablenotemark{g} &\nodata\tablenotemark{g} 
&$2.2\pm0.3$ &0.07\\
&&&1.2 &$34.0\pm6.8$\tablenotemark{i} &$106\pm21$\phn &$2.1\times2.0$
&$1600\times1500$ &$\phn-1\arcdeg$ \\
\enddata
\tablenotetext{a}{Position reported is the measured at 3.1~mm. Typical 
uncertainty in position is $0\farcs2$.}
\tablenotetext{b}{Peak intensity. The uncertainty in the
values of the flux density is $\sim 20\%$.}
\tablenotetext{c}{Integrated flux density, corrected for primary beam
response. The uncertainty in the values of the flux density is $\sim
20\%$.}
\tablenotetext{d}{Deconvolved size and P.A.\ obtained by fitting an
elliptical Gaussian.}
\tablenotetext{e}{Spectral index between 3.1 and 1.2~mm.}
\tablenotetext{f}{Total (gas+dust) circumstellar mass, obtained
assuming a dust temperature of 25~K (see $\S~4.1$).}
\tablenotetext{g}{Source not resolved.}
\tablenotetext{h}{Upper limit for non-detected sources is $4\sigma$.}
\tablenotetext{i}{Peak intensity inside the beam obtained by combining the
C and the B configurations, $1\farcs5\times1\farcs3$ at
P.A.= $7\degr$.}

\end{deluxetable}

\clearpage

\begin{deluxetable}{lccccc}
\tabletypesize{\scriptsize}
\tablewidth{0pt}
\tablecaption{Properties of the CO outflows
\label{tco}}
\tablehead{
&
\colhead{$v$\tablenotemark{a}}&
\colhead{$M$\tablenotemark{b}}&
\colhead{$P$\tablenotemark{c}}&
\colhead{$E$\tablenotemark{c}}&
\colhead{$\dot P$\tablenotemark{c}}
\\
\colhead{Lobe}&
\colhead{(km\,s$^{-1}$)}&
\colhead{($10^{-2} M_\sun$)}&
\colhead{($M_\sun$\,km\, s$^{-1}$)}&
\colhead{($10^{44}$ erg)}&
\colhead{($M_\sun\, \mbox{km s}^{-1}$\,yr$^{-1}$)}
}
\startdata 
&&&E--W OUTFLOW\phd \\
\cline{1-6}
Blue &[$-21.5, -3.5$] &7.2  &1.8 &5.7   &$6.3\times10^{-4}$\\
Red &[$+3.5, +21.5$]    &6.4  &1.8 &6.3   &$7.3\times10^{-4}$\\
Total &             &13.6\phn  &3.6 &12.0\phn  &$13.6\times10^{-4}\phn$ \\
\cline{1-6}
&&&N--S OUTFLOW\phd \\
\cline{1-6}
Blue &[$-9.5, -3.5$] &0.7 &0.1 &0.2 &$0.4\times10^{-4}$\\
Red  &[$+3.5, +9.5$]  &0.5 &0.1 &0.2 &$0.3\times10^{-4}$\\
Total &             &1.2 &0.2 &0.4 &$0.7\times10^{-4}$\\
\enddata
\tablenotetext{a}{ Range of outflow velocities with respect to the local 
standard of rest (LSR). Momenta and kinetic energies are calculated relative to
the cloud velocity, which is taken to be $v_{\rm lsr}=0$~\kms.}
\tablenotetext{b}{ The excitation temperature is 20~K. The molecular mass was
estimated following Scoville et al.\ (1986).} 
\tablenotetext{c}{ The inclination angle with respect to the plane of the sky
assumed to measure the momenta and kinetic energies is $15\arcdeg$. The CO
momentum, the kinetic energy, and the momentum rate in the outflow were
estimated following Lada (1985).} 

\end{deluxetable}

\clearpage

\begin{deluxetable}{lcccc}
\tabletypesize{\scriptsize}
\tablewidth{0pt}
\tablecaption{Column Densities of the Shocked Gas
\label{txoc}}
\tablehead{
&
\colhead{$N$(CO)}&
\colhead{$N$(CS)}&
\colhead{$N$(\metha)}
\\
\colhead{Position}&
\colhead{(cm$^{-2}$)}&
\colhead{(cm$^{-2}$)}&
\colhead{(cm$^{-2}$)}
}
\startdata
B &$1.2\times10^{16}$ &$5.9\times10^{13}$ &$1.5\times10^{15}$ \\
R &$2.8\times10^{16}$ &$7.4\times10^{13}$ &$2.0\times10^{15}$ 
\enddata

\end{deluxetable}

\clearpage

\begin{deluxetable}{lcccccc}
\tabletypesize{\scriptsize}
\tablewidth{0pt}
\tablecaption{Fractional Abundances Relative to CO
\label{tabun}}
\tablehead{
&
\multicolumn{2}{c}{IRAS 21391+5802}&&&
\multicolumn{2}{c}{L1157\tablenotemark{c}}
\\
\cline{2-3}
\cline{6-7}
\colhead{Molecule}&
\colhead{Clump B}&
\colhead{Clump R}&
\colhead{IRAS 1693$-$2422\tablenotemark{a}}&
\colhead{L1641-N\tablenotemark{b}}&
\colhead{Cloud}&
\colhead{Outflow}
}
\startdata
CS     &$5\times10^{-3}$ &$3\times10^{-3}$ &$1\times10^{-5}$ &$5\times10^{-5}$
&$3\times10^{-5}$ &$2\times10^{-3}$\\
\metha &$1\times10^{-1}$ &$7\times10^{-2}$ &$4\times10^{-5}$ &$2\times10^{-4}$
&$5\times10^{-4}$ &$0.4-2\times10^{-1}$
\enddata
\tablenotetext{a}{Blake et al.\ (1994); van Dishoeck et al.\ (1995).} 
\tablenotetext{b}{McMullin et al.\ (1994).} 
\tablenotetext{c}{Bachiller \& P\'erez Guti\'errez (1997).} 
\end{deluxetable}

\end{document}